\documentclass[prl,twocolumn,preprintnumbers,amsmath,amssymb,noshowpacs,superscriptaddress]{revtex4}
\usepackage{graphicx}% Include figure files
\usepackage{dcolumn}% Align table columns on decimal point
\usepackage{bm}% bold math

\begin{document}

\title{How perfect can graphene be?}
\author{P. Neugebauer}
\affiliation{Grenoble High Magnetic Field Laboratory, CNRS, BP 166, F-38042
Grenoble Cedex 09, France}
\author{M. Orlita}
\email{orlita@karlov.mff.cuni.cz} \affiliation{Grenoble High Magnetic Field
Laboratory, CNRS, BP 166, F-38042 Grenoble Cedex 09, France}
\affiliation{Institute of Physics, Charles University, Ke Karlovu 5, CZ-121~16
Praha 2, Czech Republic} \affiliation{Institute of Physics, v.v.i., ASCR,
Cukrovarnick\'{a} 10, CZ-162 53 Praha 6, Czech Republic}
\author{C. Faugeras}
\affiliation{Grenoble High Magnetic Field Laboratory, CNRS, BP 166, F-38042
Grenoble Cedex 09, France}
\author{A.-L. Barra}
\affiliation{Grenoble High Magnetic Field Laboratory, CNRS, BP 166, F-38042
Grenoble Cedex 09, France}
\author{M. Potemski}
\affiliation{Grenoble High Magnetic Field Laboratory, CNRS, BP 166, F-38042
Grenoble Cedex 09, France}
\date{\today}

\begin{abstract}
We have identified the cyclotron resonance response of purest graphene
ever investigated,  which can be found in nature on the surface of bulk
graphite, in form of decoupled layers from the substrate material.
Probing such flakes with Landau level spectroscopy in the THz range at very low magnetic
fields, we demonstrate a superior electronic quality of these ultra low density
layers ($n_0\approx3\times10^9$~cm$^{-2}$) expressed by the carrier mobility in
excess of 10$^7$~cm$^2$/(V.s) or scattering time of $\tau\approx20$~ps. These parameters
set new and surprisingly high limits for intrinsic properties of graphene and represent
an important challenge for further developments of current graphene technologies.
\end{abstract}

\maketitle

Fabrication of graphene structures has triggered vast research efforts
focused on the properties of two-dimensional systems with massless Dirac
fermions. Nevertheless, further progress in exploring this quantum
electrodynamics system in solid-state laboratories seems to be limited by
insufficient electronic quality of manmade structures and the crucial question
arises whether existing technologies have reached their limits or major
advances are in principle possible. The substrate, and more general any surrounding medium, has been recently
identified as a dominant source of extrinsic scattering mechanisms, which
effectively degrade the electronic quality of currently available graphene
samples~\cite{MartinNaturePhysics07,TanPRL07}. Despite significant advances in
technology, including the fabrication of suspended
specimens~\cite{BolotinPRL08,DuNN08}, the realistic limits of the scattering
time and mobility in graphene, achievable after elimination of major extrinsic
scattering sources, remain an open issue. Experiments call for higher quality
samples, which are almost certainly crucial for possible verification of
interesting predictions concerning basic phenomena of quantum electrodynamics
(e.g. Zitterbewegung) or observation of the effects of interactions between
Dirac fermions, (resulting e.g. in the appearance of the fractional quantum
Hall effect). Even more simple effects such as lifting the degeneracy of the
spin and/or pseudo-spin degree of freedom in very high magnetic fields indicate
the strong influence of the sample quality on the information that can be
deduced from experiments~\cite{AbaninPRL07,JiangPRL07II}.

\begin{figure}[b]
\scalebox{1.42}{\includegraphics*{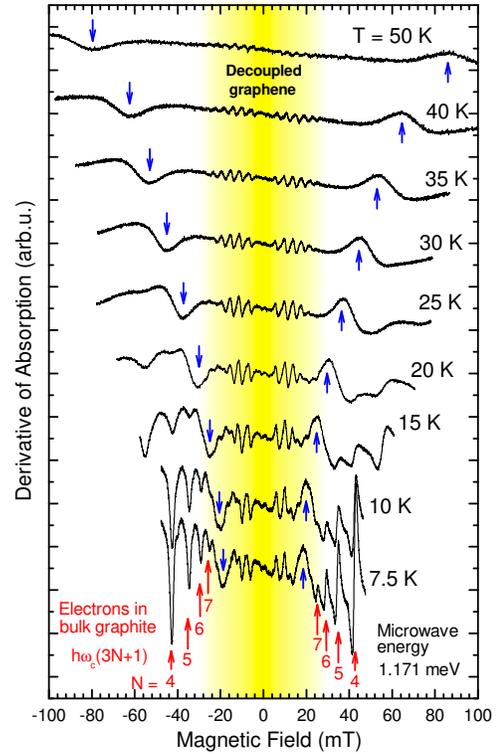}}
\caption{\label{TempDep} Magneto-absorption spectra (detected
with field modulation technique) of natural graphite specimen
taken in the temperature interval $7.5-50$~K at fixed microwave
energy $\hbar\omega=1.171$~meV. The response of low electron
concentration graphene layers decoupled from bulk graphite is seen
within the yellow highlighted area. High CR harmonics of $K$ point
electrons in bulk graphite~\cite{NozieresPR58}, with basic CR frequency $\omega_c$ corresponding to the effective mass
$m=0.054m_0$~\cite{Brandt88} are shown by red arrows. The origin of features denoted by blue arrows is discussed
in the text.}
\end{figure}

\begin{figure}
\scalebox{0.95}{\includegraphics*{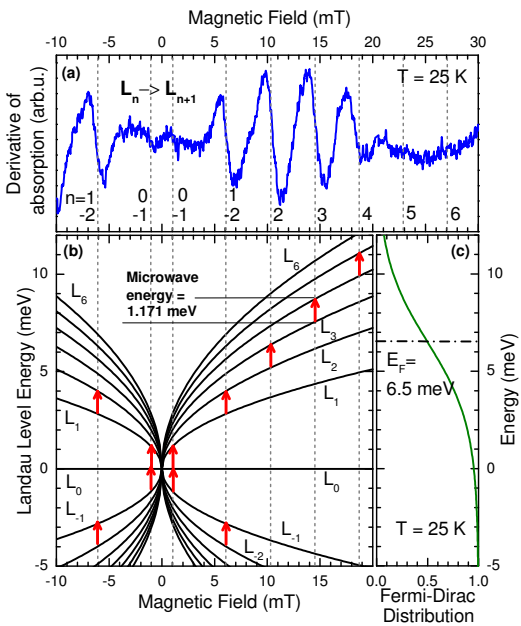}} \caption{\label{Explain}
Magneto-absorption spectrum (after removing a weak linear background seen in
the raw data of Fig.~\ref{TempDep}) measured at $T=25$~K and microwave
frequency $\hbar\omega=1.171$~meV (a) in comparison with the LL fan chart (b),
where the observed CR transitions are shown by arrows. The occupation of
individual LLs is given by the Fermi-Dirac distribution plotted in the part
(c). For simplicity, we considered only $n$-type doping with $E_F=6.5$~meV. The
dashed lines show positions of resonances assuming $\tilde{c}=1.00\times
10^6$~m.s$^{-1}$. }
\end{figure}

Recently, well-defined graphene flakes have been discovered in form of sheets,
decoupled from, but still located on the surface of bulk graphite which
naturally serves as a well-matched substrate for graphene~\cite{LiPRL09}.
In this Letter, we report on response from these flakes in microwave magneto-absorption
experiments and show that their Dirac-like electronic states are quantized into
Landau levels (LLs) in magnetic fields down to 1~mT and at elevated
temperatures up to 50~K. The deduced unprecedented quality of the studied
electronic system sets surprisingly high limits for the intrinsic scattering
time and the corresponding carrier mobility in graphene.

The cyclotron resonance has been measured in a high-frequency EPR setup in
double-pass transmission configuration, using the magnetic-field-modulation
technique. A flake of natural graphite was placed in the variable temperature
insert of the superconducting solenoid and via quasi-optics exposed to the
linearly polarized microwave radiation emitted by a Gun diode at frequencies
283.2 or 345 GHz (1.171 or 1.427 meV). The absorbed radiation has been followed
by either heterodyne detection (283.2 GHz) or by the bolometer (345 GHz). To
enhance the relatively weak response of graphene flakes, the modulation
amplitude ($\Delta B\sim0.5$~mT) had to be chosen close to the CR width, which
broadens resonances observed at lower magnetic fields. All spectra have been
corrected for the remanent field of the magnet.

Our main experimental finding is illustrated in Fig.~\ref{TempDep}. The traces
in this figure represent the magneto-absorption response of the natural
graphite specimen at different temperatures, measured as a function of the
magnetic field at fixed microwave frequency. They correspond to the derivative
of the absorption strength with respect to the magnetic field since the field
modulation technique has been applied. Strong at 7.5~K, but rapidly vanishing
with temperature lines marked with red arrows in Fig.~\ref{TempDep} can be
easily recognized as cyclotron resonance (CR) harmonics of $K$ point electrons
in bulk graphite. A possible origin of the temperature dependent transition
marked with the blue arrow will be discussed later on. The features of primary
interest, which we argue are due to decoupled graphene sheets on the graphite
surface, are seen at very low fields, within the yellow highlighted area of
Fig.~\ref{TempDep}.

The interpretation of these low-field data is schematically illustrated in
Fig.~\ref{Explain}. The observed spectral lines (Fig.~\ref{Explain}a) are
assigned to cyclotron resonance transitions between adjacent LLs ($|\Delta
n|=1$) with energies: $E_n=\mathrm{sign}(n)\tilde{c}\sqrt{2e\hbar
B|n|}=\mathrm{sign}(n)E_1\sqrt{B|n|}$~\cite{NovoselovNature05,ZhangNature05},
characteristic of massless Dirac fermions in graphene sheets with an effective
Fermi velocity $\tilde{c}$. This velocity is the only adjustable parameter
required to match the energies of the observed and calculated CR transitions. A
best match is found for $\tilde{c}=(1.00\pm0.02)\times10^6$~m.s$^{-1}$ in fair
agreement with values found in multilayer epitaxial
graphene~\cite{SadowskiPRL06,OrlitaPRL08II} or exfoliated graphene on
Si/SiO$_2$ substrate~\cite{JiangPRL07,DeaconPRB07,LiNaturePhys08}. As can be
seen from Figs.~\ref{Explain}b,c the multi-mode character of the measured
spectra is directly related to thermal distribution of carriers among different
LLs. The intensity of a given transition is proportional to the difference in
thermal occupation of the involved LLs. Roughly speaking, the strongest
transitions imply LLs in the vicinity of the Fermi level, which fixes E$_{F}$
at around 6-7~meV from the Dirac point.

\begin{figure}
\scalebox{1.2}{\includegraphics*{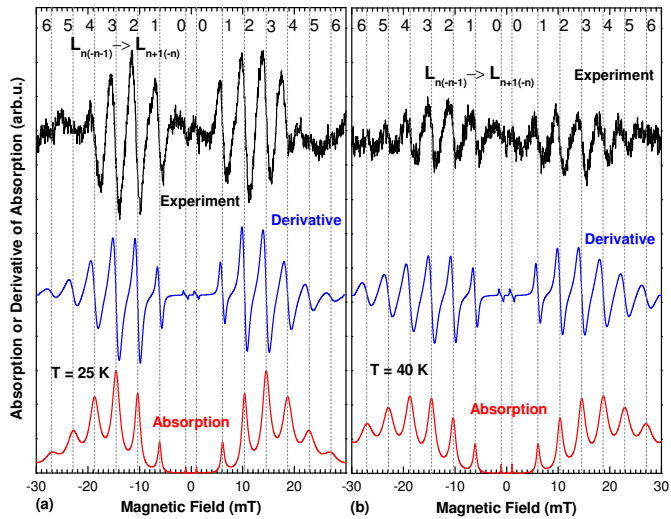}}
\caption{\label{Comparison} Background-removed magneto-absorption
spectra from experiment at $T=25$ and 40~K (black curves) in
comparison to the response at the microwave frequency
$\hbar\omega=1.171$~meV calculated for parameters
$\tilde{c}=1.00\times 10^6$~m.s$^{-1}$, $\gamma=35$~$\mu$eV and
$E_F=6.5$~meV. The dashed lines show the calculated positions of
the CR transitions.}
\end{figure}

To reproduce the experimental data, we assume the absorption strength is
proportional to the longitudinal conductivity of the system:
$$\sigma_{xx}(\omega,B)\propto
(B/\omega)\sum_{m,n}M_{m,n}\frac{f_n-f_m}{E_m-E_n-(\hbar\omega+i\gamma)},$$
where $f_n$ is occupation of the $n$-th LL and
$M_{m,n}=\alpha\delta_{|m|,|n|\pm 1}$ with $\alpha=2$ for $n$ or $m$ equal 0
otherwise $\alpha=1$~\cite{SadowskiPRL06,GusyninPRL07}. The calculated traces in
Fig.~\ref{Comparison} have been drawn taking $\gamma=35$~$\mu$eV for the line
broadening, $\tilde{c}=1.00\times10^6$~m.s$^{-1}$ and $E_F=6.5$~meV. To
directly simulate the measured traces, the derivative of the absorption with
respect to the magnetic field has been calculated taking account of the field
modulation $\Delta B=0.5$~mT used in the experiment. In spite of its
simplicity, our model is more than in a qualitative agreement with our
experimental data, see Fig.~\ref{Comparison}. The calculation fairly well
reproduces the experimental trends: Multi-mode character of the spectra, the
intensity distribution among the lines, as well as its evolution with
temperature, and allows to estimate the characteristic broadening of the CR
transitions.

Our modeling could be further improved but at the expenses of additional
complexity which we want to avoid here. Assuming magnetic-field and/or LL index
dependence of the broadening parameter $\gamma$ and taking into account the
possible fluctuation of the Fermi level within the ensemble of probed flakes
would improve the agreement between experiment and theory, particularly at low
temperatures. Comparing both, the measured and simulated traces, we are also
more confident in the spectacular observation of the CR transition (involving
the $n=0$ LL) at a magnetic field as low as 1~mT. Bearing in mind the small
value of the extracted broadening parameter one may conclude that LL
quantization should survive in studied graphene layers down to the field of
$B=(\gamma/E_1)^2\approx 1$~$\mu$T. Hence, the magnetic field of the Earth of
$\sim50$~$\mu$T is no longer negligibly small. Instead, it can open an energy
gap at the Dirac point up to $\Delta\approx0.3$~meV, depending on the sample
orientation.

\begin{figure}
\scalebox{1.2}{\includegraphics*{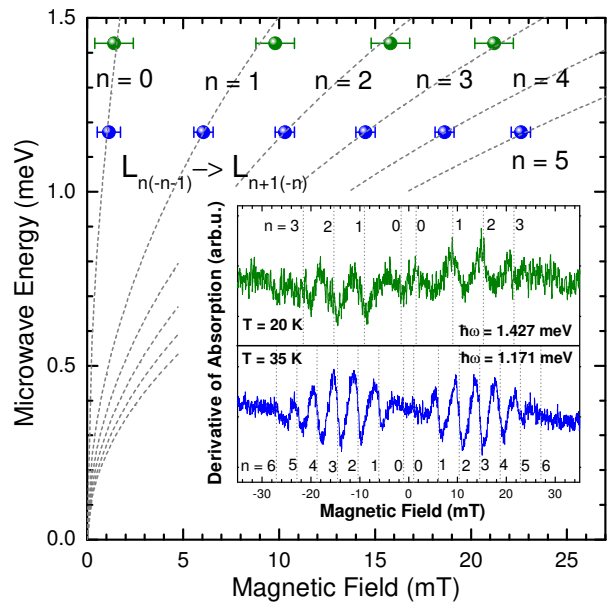}} \caption{\label{FreqDep} Positions
of CRs at microwave energies $\hbar\omega=1.171$ and 1.427~meV. Each position
represents an average value obtained at different temperatures. One
background-removed CR spectrum at each microwave frequency with a high number
of observed CRs is shown in the inset. The grey curves show positions of CRs
for $\tilde{c}=1.00\times 10^6$~m.s$^{-1}$.}
\end{figure}

To crosscheck our interpretation, we have also measured the spectra using a
different (higher) microwave energy $\hbar\omega=1.427$~meV, see
Fig.~\ref{FreqDep}. Despite the weaker sensitivity of the experimental setup at
this frequency, we can clearly identify the same set of inter-LL transitions
simply shifted to higher magnetic fields.

The finite Fermi energy $E_F\approx 6-7$~meV, corresponding to a carrier
density of $n_0\approx3\times10^9$~cm$^{-2}$, indicates that the probed layers
are in thermodynamical contact with the surrounding material, which supplies
these carriers. On the other hand, we find no signs of electrical coupling of
these graphene layers to bulk graphite. Our experiments show that any possible
energy gap opened due to this interaction at the Dirac point cannot exceed a
few hundred $\mu$eV. The absence of this gap convincingly confirms that we are
indeed dealing with decoupled graphene and not with the $H$ point of bulk
graphite, where Dirac-like fermions are also present~\cite{OrlitaPRL08} but a
(pseudo)gap of a few meV is expected~\cite{GruneisPRB08} and indeed
observed~\cite{ToyPRB77}. The temperature evolution of the measured spectra is
another important indication allowing us to discriminate between the graphene
and bulk graphite contributions. No (or very weak) temperature broadening of CR
transitions is expected for graphene
\cite{MorozovPRL08,BolotinPRL08,OrlitaPRL08II}, whereas the response of bulk
graphite should follow the relatively strong decrease of the carrier scattering
time, expressed by the average mobility, which reaches up to
$10^6$~cm$^2$/(V.s) at low temperatures, but falls down by one order of
magnitude at $T\approx50$~K~\cite{Brandt88}. Indeed, this behaviour is observed
in Fig.~\ref{TempDep}. Whereas the CR harmonics of $K$ point electrons in bulk
graphite~\cite{NozieresPR58} seen in the spectra at $|B|\gtrsim 20$~mT, disappear very rapidly
upon increasing $T$, the graphene-like features survive and their intensity is simply following the vanishing difference in the occupation
between the adjacent LLs. It is worth noticing that the graphene-like signal,
although always substantially weaker then the response from bulk graphite, has
been observed for a number of different specimens of natural graphite.
Mechanical scratching of the sample surface and fast thermal cooling of the sample has been found to enhance the signal from
decoupled graphene in comparison with bulk graphite, likely helping in
detaching graphene sheets from the graphite crystal.

Since the well-defined LL quantization in our graphene flakes is observed down
to $|B|=1$~mT, see Fig.~\ref{Explain}, we obtain via the semi-classical LL
quantization condition $\mu B>1$ the carrier mobility
$\mu>10^7$~cm$^{2}$/(V.s), almost two orders of magnitude higher in comparison
with suspended~\cite{TanPRL07,BolotinPRL08,DuNN08} or epitaxial
graphene~\cite{OrlitaPRL08II}. The LL broadening $\gamma$, obtained via
comparison of our experiment with the simulated traces, allows us to estimate
the scattering time $\tau\approx20$~ps ($\tau=\hbar/\gamma$), which
significantly exceeds those reported in any kind of manmade graphene samples,
see e.g. Refs.~\cite{TanPRL07,BolotinPRL08}, and gives an independent
estimation for the mobility $\mu=e\tau\tilde{c}^2/E_F\approx
3\times10^7$~cm$^{2}$/(V.s) in good agreement with the estimate above. Even
though we cannot verify this estimate by a direct electrical measurement, a
near correspondence of the scattering time derived from CR measurements and
transport scattering time was recently verified on samples with a significantly
lower mobility~\cite{JiangPRL07}. Moreover, the estimated mobility should not
decrease with temperature, as no broadening of CRs is observed up to $T=50$~K,
when CR intensities become comparable with the noise. This extremely high value
of mobility combines two effects: the long scattering time $\tau$ and a very
small effective mass
$m=E_F/\tilde{c}^2\approx2\times10^{-4}m_0E_F[\mathrm{meV}]$. Remarkably, the
same scattering time in a moderate density sample ($n_0=10^{11}$~cm$^{-2}$),
would imply the mobility still remaining high, around
$\mu\approx5\times10^6$~cm$^{2}$/(V.s), and comparable to best mobilities of
two-dimensional electron gas in GaAs structures at these densities.

The model we used here to describe the magneto-absorption response of Dirac
fermions in graphene is amazingly simple, based on a one-particle
approximation, and it is perhaps surprising that it is so well applicable to
simulate the experimental data, particularly in context of the outstanding
quality of the electronic system studied. At first sight, the observation of
collective excitations, due to, e.g., magneto-plasmons could be expected in our
experiments and only preliminary theoretical work addresses the surprising
approximate validity of Kohn's theorem in graphene~\cite{BychkovPRB08,Ando}.
Also size-confined magneto-plasmons are apparent in the microwave-absorption
spectra of two-dimensional gas of massive electrons~\cite{AllenPRL77}, but
apparently not seen in our experiments on graphene. This perhaps points out the
qualitative difference in the plasmon behaviour in systems with quadratic and
linear dispersion relations~\cite{HwangPRB07,PoliniCM08}. On the other hand, we
speculate that the transition, marked with blue arrows in Fig.~\ref{TempDep}
has a character of a collective excitation. It is characterized by a
sensitivity to the magnetic field and by softening with temperature. The energy
of this excitation can be deduced to scale roughly as $B/T$, and this points
towards its magnetic origin. For example, $B/T$ energy scaling is typical of a
ferromagnetic resonance, see e.g. Ref.~\onlinecite{LiuJPCM06}. Nevertheless,
the origin of the spectral feature marked by blue arrows in Fig.~\ref{TempDep}
remains an intriguing puzzle. We see this feature repeatedly on different
samples, but at slightly different energies, each time however
characteristically evolving with temperature. With its apparent magnetic
characteristics one should seek its origin in impurities~\cite{UchoaPRL08},
structural defects~\cite{YazyevPRB07} or edge states~\cite{YangPRL08} in
graphene, but possibly also on the surface of bulk
graphite~\cite{CervenkaNP09}.

To conclude, graphene layers decoupled from bulk graphite have been probed in
cyclotron resonance experiment, which offers an unambiguous evidence of
extremely high carrier mobility in graphene exceeding $10^7$~cm$^2$/(V.s). This
measurement significantly shifts the limits of intrinsic mobility in
graphene~\cite{MorozovPRL08} and poses a quest for further development in the
technology of graphene fabrication. Graphene samples with mobilities comparable
to the best GaAs samples~\cite{HwangPRB08II} thus seem to be achievable.

\begin{acknowledgments}
P.N. and M.O. contributed equally to this work. Part of this work has been
supported by EuroMagNET II under the EU contract, by the French-Czech Project
Barrande No.~19535NF, by contract ANR-06-NANO-019 and by projects MSM0021620834
and KAN400100652.
\end{acknowledgments}

%\bibliography{PRLVersion}

\end{document}